\documentclass{eas}
\usepackage{graphicx}
\newcommand{\micron}{$\mu$m}
\newcommand{\etl}{{\em et al.\/}} 
\newcommand{\aap}{A\&A}
\newcommand{\apj}{ApJ}
\newcommand{\apjl}{ApJ Letters}
\newcommand{\apjs}{ApJ Supp}
\newcommand{\apss}{ApSpS}
\newcommand{\mnras}{MNRAS}
\newcommand{\pasp}{PASP}
\newcommand{\qjras}{QJRAS}
\begin{document}

\title{On Predicting the Polarization of Low Frequency Emission by
Diffuse Interstellar Dust}
\runningtitle{P. G. Martin: On the Low Frequency Polarization of Cirrus}
\author{P. G. Martin}\address{Canadian Institute for Theoretical
Astrophysics, University of Toronto;\\
\email{pgmartin@cita.utoronto.ca}}
\begin{abstract}
Several of the current and next-generation cosmic microwave background
(CMB) experiments have polarimetric capability, promising to add to
the finesse of precision cosmology.  One of the contaminating Galactic
foregrounds is thermal emission by dust.  Since optical interstellar
polarization is commonly seen, from differential extinction by aligned
aspherical dust particles, it is expected that thermal emission from
these grains will be polarized.  Indeed, in the Galactic plane and in
dark (molecular) clouds, dust emission in the infrared and
submillimetre has been measured to be polarized.  It seems likely that
the faint diffuse cirrus emission, of more relevance to CMB
experiments, will be polarized too.  We discuss how well the amount of
polarization of this component can be predicted, making use of what is
known about optical (and infrared and ultraviolet) interstellar
polarization and extinction.  Some constraints on the alignment of the
carrier of the dust-correlated anomalous microwave emission can be
made as well.
\end{abstract}
\maketitle
\section{Introduction: CMB Polarization and Cirrus Contamination}\label{int}

\begin{figure}
\includegraphics[width=5cm]{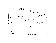}
\qquad
\includegraphics[width=5cm]{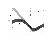}
\caption{Left: Foreground components contaminating the CMB at two
representative latitudes. From www.planck.fr/heading136.html Giard \&
Lagache.  Right: Alternative presentation from de Oliveira-Costa \etl\
(\cite{deo04}) showing the anomalous emission as well.
}
\label{contam}
\end{figure}

Several of the current and next-generation cosmic microwave background
(CMB) experiments have the capability of measuring linear
polarization, a complementary probe for precision cosmology (e.g.,
Laureijs \cite{lau06}; Polenta \etl\ \cite{pol05}; Beno\^it \etl\
\cite{ben04}; Prunet \cite{pru06}).
Among these experiments are Archeops (Beno\^it \etl\ \cite{ben04}),
BOOMERanG (B2K in 2003; Polenta \etl\ \cite{pol05}; Netterfiled
\cite{net06}), and the Planck Surveyor (Laureijs \cite{lau06}; Prunet
\cite{pru06}; Lammare \etl\ \cite{lam03}).
All-sky surveys by IRAS from 12 to 100~\micron\ revealed faint diffuse
emission (``cirrus'') everywhere, even at high Galactic latitudes
(Low \etl\ \cite{low84}).  
One of the diffuse foregrounds contaminating the CMB signal is thermal
emission by this diffuse interstellar dust (e.g., de Oliveira-Costa
\etl\ \cite{deo04}; Tucci \etl\ \cite{tuc05}), expected to correspond
most closely spatially to the 100~\micron\ emission (\S~\ref{comp}).
This becomes the dominant foreground above about 100 GHz ($\lambda <
3$~mm $=3000$~\micron), and as the CMB fades toward higher
frequencies, the rising dust spectrum becomes the dominant signal
(Fig.~\ref{contam}).
Where polarization is concerned, some further caution is needed in
defining where the dust dominates.  The frequency above which the
polarized dust intensity exceeds the polarized synchrotron background
is probably somewhat higher because the synchrotron component is
likely to have a higher degree of polarization.  Of course, it is the
spatial fluctuations, which will vary from component to component,
that are in the end important to CMB contamination.
Multifrequency observations through the transitional range are
therefore a prerequisite for separating these signals (component
separation, not a specific topic of this paper; see e.g.,
Jaffe \etl\ \cite{jaf04}).

Some CMB experiments like B2K minimize cirrus contamination by
concentrating measurements in regions of low foreground column density
(Netterfield \cite{net06}).
One strategy for cirrus mitigation could therefore be to mask out
regions of bright cirrus.  But only 20\% of the sky has an HI column
density below $10^{20}$~cm${-2}$ (Miville-Desch\^enes, private
communication) and even that produces a non-negligible foreground
($\sim 1$~MJy~sr${-1}$ at 100~\micron).  The precision sought by
instruments like the Planck Surveyor depends on broad sky coverage
(Prunet \cite{pru06}), and of course Planck will in any case produce
data for the entire sky.
Thus, as an alternative strategy, one needs to measure the properties
of cirrus at higher frequencies where the CMB is not important, and
then extrapolate to lower frequencies where one does have to address
component separation.  To inform/confirm the extrapolation, the cirrus
properties at the lower frequencies can be assessed empirically in
regions of brighter cirrus.  The spectral properties of cirrus are of
course an important diagnostic of the interstellar dust.

Since optical polarization is commonly seen, from differential
extinction by aligned aspherical dust particles, it is expected that
thermal emission from these grains will be polarized (Stein
\cite{ste67}).
Indeed, in the Galactic plane and in dark (molecular) clouds, dust
emission in the infrared and submillimetre has been measured to be
polarized (Vaillancourt \cite{vai06}; Hildebrand \etl\ \cite{hil00};
Vaillancourt \etl\ \cite{val03}).  In these dense regions, the
interpretation of the polarization is complicated by many issues: beam
dilution, distortions in the magnetic field topology, changes in the
degree of alignment within clouds, grain evolution, and a range of
grain temperatures and optical depth which affect which grains
dominate the emission and polarization in various parts of the
infrared to submillimetre spectrum.  These are not considered here.

It seems likely that the faint diffuse cirrus emission will be
polarized too, and this will not be affected by the above-mentioned
complications.  Diffuse dust polarization is of more relevance to CMB
experiments.  Polarimetric observations in the transitional high
frequency range (see Table~\ref{angular}) were made by Archeops at 353
GHz (Beno\^it \etl\ \cite{ben04}) and B2K and MAXIPOL at lower
frequencies (Polenta \etl\ \cite{pol05}; Johnson \etl\ \cite{jon03}),
precursors to Planck HFI.
Archeops has reported detecting the dust polarization (\S~ref{sobs}).
From the point of view of best examining the dust polarization,
Planck has no polarimetric capability at
545 and 857~GHz where the dust polarized intensity would be the
strongest.  However, balloon-borne experiments like PILOT (Bernard
\cite{ber05}) and BLAST-pol will map the diffuse polarization with
great sensitivity at higher frequencies.

This paper attempts to quantify the degree to which the diffuse dust
emission is polarized, a contribution toward both component separation
and dust physics.

\begin{table}[!b] 
\caption{Angular Resolution$^1$ of High-Frequency CMB Polarization Experiments}
\begin{center}
\begin{tabular}{lccccccccc}
\hline \\
$\nu$ (GHz) &      100  &  143 & 150  & 217  & 240  & 345 & 353  & 545 & 1250 \\
$\lambda$ (\micron) & 3300 & 2100 & 2000 & 1380 & 1250 & 870 & 850  & 550 & 240 \\
\hline \\
Archeops &&&&&&& 13 \\ 
BOOMERanG 2K &&& 10 && 7 & 7 \\  
MAXIPOL & & 10 && & 10 \\
PILOT & &&&&&&& 2.3 & 1 \\
%
Planck HFI & 9.2 &7.1 && 5 &&& 5\\
\hline \\
\end{tabular}
\end{center}
{$^1$ FWHM in arc minutes} \\
\label{angular}
\end{table}

WMAP (Kogut \etl\ \cite{kog03}) and Planck have polarimetric capability at lower
frequencies, down to 23 GHz and 30 GHz respectively, which can be used
to examine the polarization of two other CMB foregrounds, synchrotron
emission and the dust-correlated anomalous microwave (hereafter simply
``anomalous'') emission (Draine \& Lazarian \cite{dra98}; Finkbeiner
\etl\ \cite{fin99}; Lazarian \& Finkbeiner \cite{laz03}; Finkbeiner
\cite{fin04}; Finkbeiner \etl\ \cite{finl04}; de Oliveira-Costa \etl\
\cite{deo04}; Davies \cite{davies06}; Davis \cite{davis06}).
Some constraints on the alignment of a dust carrier of the latter
emission, which would be diagnostic of the emission mechanism, are
made here as well (\S~\ref{anomalous}).

\subsection{A Legacy from Optical Polarization} \label{legacy}

\begin{figure}[!b]
\includegraphics[width=8cm]{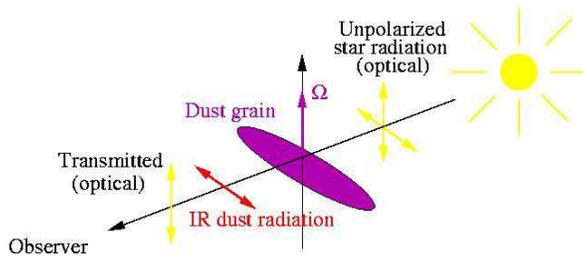}
\caption{The relationship of the polarization of emitted and
transmitted radiation to the mean grain profile, aligned with respect
to the Galactic magnetic field.  In the far-infrared and
submillimetre, there is more radiation emitted with polarization
E-vector along the long axis.  This is also the axis along which there
is more extinction, and so the E-vector of the net transmitted
radiation is orthogonal to this axis, and to the emitted polarization and
the magnetic field.  Graphic from Ponthieu \& Lagache at
www.planck.fr/article263.html.
}
\label{dustrad}
\end{figure}

Both optical interstellar polarization and polarization of thermal
emission depend on aspherical (flattened or elongated) grains that are
aligned.
The orientation of the electric vector of the emitted radiation is
along the long axis of the mean grain profile projected on the plane
of the sky, while that of the optical polarization, being of
transmitted light suffering differential extinction, is orthogonal to
this.
%
%
Particular ingredients in alignment theories are that grains are
rapidly spinning about their axis of largest moment of inertia and
that this is on average aligned with the Galactic magnetic field, so
that consistent with observations the electric vector of optical
polarization is along the direction of the projected magnetic field,
while it is perpendicular for emission (see Fig.~\ref{dustrad}).
There are several potentially viable alignment mechanisms as discussed
elsewhere (Lazarian \& Finkbeiner \cite{laz03}; Lazarian \& Yan
\cite{laz04}; Roberge \cite{rob04}; Vaillancourt \cite{vai06}) and in
the Appendix.
Nevertheless, it is still difficult to predict the degree of alignment
with certainty {\it ab initio}, or as will become clear, gauge other
important factors, and so a focus of this paper is to assess how well
the degree of polarization of the diffuse cirrus component can be
predicted semi-empirically.

To predict the degree of polarization of the dust-related CMB
foreground(s), we draw on what can be deduced about alignment from
optical (and infrared and ultraviolet) interstellar polarization and
extinction, concentrating on diffuse dust associated with atomic
gas. Still, it should be cautioned that the optical measurements are
by necessity largely for column densities with $E_{B-V} > 0.1$ and so
$N_H > 6 \times 10^{20}$~cm$^{-2}$.
This corresponds to lines of sight with 100-\micron\ cirrus brightness
$I_{100} > 6$~MJy~sr$^{-1}$.  While there will be many such directions
in which to examine the polarization of the low-frequency foreground
cirrus with forthcoming experiments, for CMB applications one needs to
work at even lower column densities too.
Another issue for any intercomparison of optical polarization and the
polarization of low-frequency dust emission is that the Galaxy is
optically thin for the latter.  Therefore, we see the whole Galaxy, or
right out of it, unlike probes with stars which rely on differential
extinction along that path.  But at high latitude, most relevant to
the CMB, the effective paths might not be too dissimilar if the stars
used were sufficiently distant.

A lot of attention goes into the degree of polarization $p = P/I$, the
ratio of the polarized intensity $P$ to the total emission $I$.
Non-polarizing grains (or other emission components entirely) can
contribute to $I$, potentially confusing the interpretation of any
spectral dependence of $p$.  We therefore also emphasize the
importance of examining $P$ and its spectral dependence directly, as a
diagnostic of the aligned grains.
$P$ is of course what is relevant to the calculation of the power
spectrum of the $E$ and $B$ modes of the polarized intensity.  
We do not attempt to predict these (Prunet \etl\ \cite{pru98}; Tucci
\etl\ \cite{tuc05}) since they depend in addition on the spatial
variation of the degree and orientation of the alignment, whose
statistical properties are not known (i.e., the statistics of $P$ are
different than the statistics of $p_{em}I$).  However, the dust
contribution to the power spectrum should scale with the same
frequency dependence as $P^2$ (from dust), unless hoped-for ``spatial
-- electromagnetic frequency decoupling'' breaks down, i.e., there
were different populations of grains with different alignment
statistics whose relative contributions to $P$ (and $I$) changed with
frequency.
A $TE$ measurement of the dust emission at 353~GHz by Archeops
(Ponthieu \etl\ \cite{pon05}) provides a basis from which the
potential contamination at lower frequencies can be estimated more
directly.

\section{Polarized Emission} \label{emission}

The polarization of thermal emission by dust was first noted and
estimated by Stein (\cite{ste67}).  Hildebrand (\cite{hil88}) has
since made more detailed calculations regarding infrared polarization
and
Hildebrand \& Dragovan (\cite{hil95}) have analyzed far-infrared
observations.
%
We begin our estimate on the same basis, from electromagnetic
scattering calculations of the cross-section for emission (equal to
that for absorption),$C_{em}$, for single grains (for much of the rest
of this section we suppress the subscript $em$).
What is often computed for a single grain (e.g., Martin \cite{mar75})
is an efficiency factor $Q = C/A$, where $A$ is some measure of the
area of the grain (whether projected or actual does not matter
here). (Note that this is not Stokes $Q$.)  But first we recall one of
the challenges, that there are many different grain components
potentially involved.

\subsection{Grain Components: Lessons from IRAS and ISO}\label{comp}

\begin{figure}
\includegraphics[width=5cm] {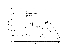}
\qquad
\includegraphics[width=5cm] {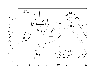}
\caption{Spectral components of high Galactic latitude cirrus in the
infrared and implications for grain sizes. Left: D\'esert \etl\
(\cite{des90}).  Right: Li \& Draine (\cite{lid01}). See text.}
\label{desert}
\end{figure}

Interstellar extinction suggests a range of grain sizes and
compositions (\S~\ref{extinction}) and the infrared spectrum of cirrus
in the range 3 to 1000~\micron\ is interpreted as the sum of several
components too (Fig.~\ref{desert}).

Near 100~\micron\ and longer, the range of interest here, the emission
is thermal emission by ``big'' grains (size $\sim 0.1$~\micron) in
equilibrium with the local interstellar radiation field.


The mid-infrared emission at 60 and 25~\micron\ is in excess of what
can be attributed to the big grains, and is interpreted as
non-equilibrium emission by grains that are very much smaller (VSGs:
0.007~\micron\  $ = 70$~\AA\ $ = 7$~nm), such that a single absorbed
ultraviolet photon raises the temperature momentarily above the
equilibrium value.

The 12~\micron\ and shorter emission has many spectral features
indicative of polycyclic aromatic hydrocarbons (PAHs).  This can be
thought of as either non-equilibrium emission by tiny grains/PAHs
(1~nm) or the result of internal conversion in large molecules (see
e.g., Draine \& Li \cite{dra01}).

All of these smaller grain components of course radiate at longer
wavelengths too.  But because the big grains contain most of the mass,
it is they that dominate in the submillimetre.

Tiny grains (including PAHs) also spin rapidly and emit microwave
radiation which could be another foreground contaminant of the
polarized CMB, if aligned.

\subsection{Submillimetre Spectrum: Total Intensity from an Ensemble of Grains}\label{stokesi}

The emitted total intensity $I$ summed over grain populations
with column density $N_g$ and temperature $T_g$, and acknowledging all
``other'' components (CMB, free-free, synchrotron, anomalous emission,
point sources, etc.) is

\begin{equation}
I_{\nu} = \Sigma_g\ B_{\nu}(T) N A Q_{\nu}(T) = \Sigma_g\ B_{\nu}(T) N
V {C_{\nu}(T)\over V} +I_{other}.
\label{emit}
\end{equation}

\noindent
Here $C$ is the mean of the two effective cross sections for
orientations of the electric vector parallel and perpendicular to the
long axis of the projected mean profile of the aligned grains (``par''
and ``per'' below in equation \ref{pemit}).
Normally $C$ depends on the grain size (characterized by some
dimension $a$) but in the millimetre to submillimetre range of
interest, $a$ is much smaller than the wavelength $\lambda$ and the
absorption cross sections per unit volume $V$, $C/V = Q/a$, is
independent of $a$.  The contributions to $I$ (and $P$) are therefore
weighted by the volume; in the diffuse interstellar medium, large
grains carry most of the volume (or mass).
It is often a good approximation in this limit that $C/V = Q/a =
\alpha \nu^\beta $, where $\beta$ is the spectral index of the
emissivity (e.g., $\beta \sim 2$ for many materials) and $\alpha$ is
the scaled opacity (with the basic frequency dependence removed), that
varies from material to material.  The thermal emission spectrum of
each separate grain component is then a modified greybody emission,
$\sim B_{\nu}(T) \nu^\beta$, with the appropriate $T$ and $\beta$.

If there are multiple grain components with different properties
(including alignment), then the contributions to $I$ (and $P$) are
weighted by the volume, $\alpha$, and the Planck function at the
appropriate temperatures, potentially making $I$ deviate from a simple
modified greybody with a frequency-independent $\beta$; nor would
$\beta$ be necessarily representative (diagnostic) of any of the grain
materials.  Spectral changes would be most pronounced near the peak of
any of the Planck functions, because of changing relative weights.

The wider the frequency range, the less likely it is that the opacity can be
described by a single $\beta$.
Changes in $\beta$ with frequency over the range of interest are seen
in laboratory experiments on amorphous silicates (Boudet \etl\
\cite{bou05}).
In fitting the spectrum, this provides an alternative to invoking an
extra diffuse cold ($< 10$~K) dust component (Finkbeiner \etl\
\cite{fin99}; Bourdin \etl\ \cite{bou02}); although this provides a
good fit and interpolation formula, one must be wary about taking the
model literally, since dust that cold in the local diffuse ISM seems
unphysical,
Also there is the possibility, often ignored, that both $\alpha$ and
$\beta$ (thus the opacity) could be temperature dependent, again
supported by the same laboratory measurements on amorphous materials,
where the effects are attributed to intrinsic processes (two level
systems, disordered charge distribution) in a single amorphous
(silicate) material.
This might also help explain why $\beta$ appears to change with $T$
(Boudet \etl\ \cite{bou05}; Dupac \etl\ \cite{dup03}), though here it
can be noted that this is for $T$ higher than in regions of relevance
to diffuse CMB foregrounds.
As discussed in \S~\ref{spectp}, an important implication might be
that silicates provide a more dominant contribution to the emission at
lower frequencies (\S~\ref{dilute}), of relevance to polarized
emission (and its detailed frequency dependence) since, among grain
components, large silicates at least are aligned
(\S~\ref{polarization}).

\subsection{Polarized Intensity}

The emitted polarized intensity $P$ (Stokes $Q$ in the
appropriately-rotated coordinate system) summed over all (aligned)
grains, and including a shorter list of ``other'' polarized
components, is similar:

\begin{equation}
P_{em} = - \Sigma_g\ B_{\nu}(T_g) N_g {1\over 2}(C_{par} - C_{per}) +
P_{other}.
\label{pemit}
\end{equation}

\noindent
Since $C_{par} > C_{per}$, the emitted light is polarized with
electric vector parallel to the long axis of the mean grain profile
projected on the plane of the sky.  The minus sign invokes a
convention, a reminder of the orthogonal orientation of the electric
vector for emission relative to that of the transmitted light (see the
sign convention in equation \ref{pot}).

\subsection{Emission Cross Sections for ``Small'' Particles}

In the relevant limit in which $a$ is much smaller than the wavelength
$\lambda$ simple analytical formulae can often be used.
These are particularly straightforward to calculate for spheroids (or
ellipsoids for that matter) that are homogeneous.  In this limit there
are only two principal cross sections to evaluate, corresponding to
electric vector orientation parallel and perpendicular to the symmetry
axis:

\begin{equation}
{ {Q_{\parallel,\perp}} \over a } = { {C_{\parallel,\perp}} \over V} =
- {2 \pi \nu \over c } {\rm Im} ( { m^2 -1 \over
L_{\parallel,\perp}(m^2 -1) + 1 } ).
\label{qspheroid}
\end{equation}

\noindent
The principal cross sections are analytic functions of the complex
refractive index $m$ (hence dependent on grain material) and the
``polarizability'' factors $L$ which depend on the shape and axial
ratio (e.g., Martin \cite{mar75}).

Interstellar grains are spinning and the largest amount of polarized
emission $P$ would arise for what we will call perfect spinning
alignment (PSA), corresponding to rotation about the grain axis of
largest moment of inertia, with all spin axes aligned along a common
axis in space (the interstellar magnetic field) which in turn is in
the plane of the sky.

For spheroids with perfect spinning alignment (PSA, as opposed to
picket fence alignment, PF) we have
$Q_{par,per} = Q_{\perp,\parallel}$ for oblate shapes and
$Q_{par,per} = (Q_{\parallel,\perp} + Q_{\perp})/2$ for prolate
shapes.

\subsection{Degree of Polarization of Aligned Grains $pa_{em}$}\label{palign}

For a single grain population (size, composition, and orientation,
leading to the same temperature $T_g$), the dependences on $N_g V$ and
$B_{\nu}(T_g)$ would cancel out in the ratio $P/I$ and so

\begin{equation}
pa_{em} = - {Q_{par} - Q_{per}\over Q_{par}+ Q_{per}} .
\label{pem}
\end{equation}
 
\noindent
In this simple situation, the degree of polarization is a property
characteristic of the grains (composition, shape, alignment), but not
their column density.

\begin{figure}
\includegraphics[width=5cm,angle=90] {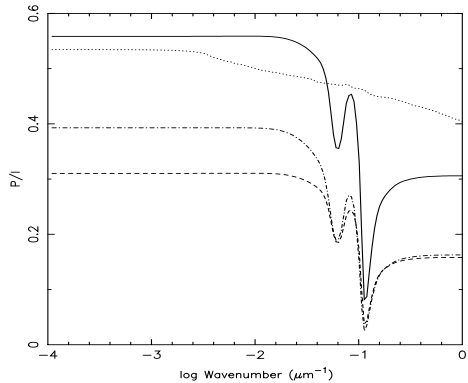}
\caption{$pa_{em}$ for perfect spinning alignment, as a function of
shape and composition.
Silicate: oblate with axial ratio 1.4 (dash) and 2 (solid); prolate
with axial ratio 2 (dash-dot).  Amorphous carbon: oblate with axial
ratio 2 (dots).}
\label{figpem}
\end{figure}

For $m$ appropriate to ``astronomical silicate'' (Draine \& Lee
\cite{dra84}; Rouleau \& Martin \cite{rou91}), Figure~\ref{figpem}
illustrates $pa_{em}$ calculated for several axial ratios.
Also shown is one curve for an amorphous carbon material (AC1; Rouleau
\& Martin \cite{roul91}).


There are several things to note.  
The degree of polarization is potentially very large, even from
particles that are only mildly aspherical (axial ratio $< 2$).
The value of $pa_{em}$ depends on the shape, axial ratio, and degree of
alignment, and so by itself is not diagnostic of the specific grain
material.
Even though $Q/a$ has a strong spectral dependence as described above,
this too cancels out in $pa_{em}$.  There can be a slight frequency
dependence of $pa_{em}$ across the range of interest, due to changes in
$m$.

In contrast, the spectral dependence of the observable $P$ does retain
the $\nu^\beta$ dependence.  The wavelength dependence of optical
polarization indicates that only large grains are preferentially
aligned, which reinforces their dominance in the volume-weighted $P$
and makes it more likely that the spectral dependence of $P$ could be
a useful diagnostic of the material of the aligned grains.
For example, if over the range $\sim 100$ -- 1500~GHz interstellar
amorphous silicates had spectral index variations as measured by
Boudet \etl\ (\cite{bou05}) or inferred by Li \& Draine (\cite{lid01})
and these were the aligned grains (silicates certainly are aligned;
see \S~\ref{polarization}), then this spectral behaviour would be
imprinted directly on $P$ with little effect on $pa_{em}$ (see the
illustration in \S~\ref{spectp}).


Non-aligned grains and $I_{other}$ dilute the degree of polarization
that would be produced by the aligned grains.  The spectral dependence
of $P$ would not change, but the interpretation of $p_{em}$ would
become more difficult.

More generally still, spectral dependence in the net $p_{em}$ can be
introduced by frequency-dependent relative weighting in $P$ and/or
$I$, from different degrees of alignment, and/or temperatures $T_g$
and opacities (many models implicitly imply these effects, e.g.,
Finkbeiner \etl\ \cite{fin99}; Bourdin \etl\ \cite{bou02}; Li \&
Draine \cite{lid01}), and from the contributions $P_{other}$ and
$I_{other}$.

\subsection{Challenges}

As mentioned above, there is a wide range of grain sizes and
compositions.  In order to predict the polarized emission, we need to
know which of these grains are aligned and how well.  This is wrapped
up with grain shape too: how flattened/elongated are the particles?
In summary, $p_{em}$ (and $P$) depends sensitively on axial ratio,
grain orientation, and composition, none of which are well known.

\section{Optical (and Infrared and Ultraviolet) Interstellar 
Polarization} \label{opti}

Fortunately, though the challenges can be met; the appropriate
combinations of these unknowns can be constrained using interstellar
polarization in the optical.  First we reinforce one of the
challenges, the multiple components suggested by interstellar
extinction curve, $\tau_{ex}(\nu)$.

\subsection{Grain Components: Lessons from the Extinction Curve}\label{extinction}

\begin{figure}[!b]
\includegraphics[width=5.5cm] {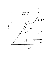}
\qquad
\includegraphics[width=6.5cm] {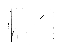}
\caption{Frequency dependence of interstellar extinction. Left:
Extinction curves in the optical and ultraviolet showing a range of
behaviour as a functon of $R_V$ (Cardelli \etl\ \cite{car89}); in the
diffuse medium $R_V \sim 3.1$.  The sole spectral feature is the
prominent ``bump'' at 2175~\AA.
%
Right: In the infrared there is a power-law decrease, plus a
distinctive 10-\micron\ silicate feature (Martin \& Whittet \cite{mar90}).
}
\label{ccmsil}
\end{figure}

Interstellar extinction is caused by scattering plus absorption by
grains comparable in size to the wavelength.  The dependence of the
cross sections $C_{ex}$ on size and frequency cannot be obtained by
the small particle approximation described above and require instead
complex electromagnetic scattering calculations such as by Mie theory.
The grain models are used to interpret observations of the
interstellar extinction curve (Fig.~\ref{ccmsil}, left).


One basic conclusion is that the continued rise in extinction into the
ultraviolet requires smaller and smaller grains, a range already known
prior to the interpretation of the infrared emission.
There is a well-characterized "bump" at 2175~\AA\ thought to arise in
carbonaceous grains.  This is another indication of separate grain
components.  In the infrared at 10~\micron\ (Fig.~\ref{ccmsil},
right), there is a prominent silicate absorption feature (Si-O
stretch), the strength and shape of which requires that most of the
available Si is depleted in amorphous silicate grains.  The
near-infrared extinction follows a power law with index near 2 (Martin
\& Whittet \cite{mar90}) which can be modeled as the scattering by
large grains comparable in size to the wavelength.

In summary, grains come in many sizes (perhaps a function of
composition).  This raises many questions.  Which grains produce the
extinction in the optical and ultraviolet?  Which grains produce the
submillimetre emission?  Which grains polarize in the optical and
ultraviolet?  Does this result in significant submillimetre
polarization?

\subsection{Lessons from the Polarization Curve}\label{polarization}

\begin{figure}[!b]
\includegraphics[width=6cm] {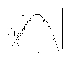}
\qquad
\includegraphics[width=5.9cm] {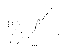}
\caption{Frequency dependence of interstellar polarization.  Left:
Polarization curve from the near-infrared to the ultraviolet,
normalized to the maximum polarization in the visual (Martin \etl\
\cite{mar99}).  There is no strong ultraviolet polarization ``bump''
as in the extinction curve.
%
Right: In the infrared there is a power-law decrease, plus a
distinctive polarized 10-\micron\ silicate feature (Martin \& Whittet
\cite{mar90}).
%
In lines of sight to an embedded source, such as this toward the
Becklin-Neugebauer object in OMC 1, there is polarization at
3.1\micron\ ice band, interpreted as a thin frost on the aligned
silicates.
}
\label{polcurve}
\end{figure}

At least some of the grains are aspherical and aligned, so that in the
plane of the sky the ensemble average grain profile is elongated,
empirically with the long axis of this profile oriented perpendicular
to the direction of the magnetic field $B$.
Differential extinction, according to the orientation of the electric
vector $E$ with respect to the mean projected grain profile, produces
a net polarization of transmitted light.  Since there is greater
extinction for $E$ parallel to long axis, then $E$ of the transmitted
radiation is parallel to short axis, hence parallel to
(Fig.~\ref{dustrad}).

Empirically, the ``polarization curve'', the frequency dependent
degree of polarization $p_{ex}(\nu)$, increases in the infrared to a
peak in the optical and then declines into the ultraviolet
(Fig.~\ref{polcurve}, left).  Computations for single sizes indeed
show a decline in polarization before the peak is reached in
extinction (Roger \& Martin \cite{rog79}).  But quantitatively this
plus the continued rise of extinction into the ultraviolet, which
implies smaller and smaller particles, plus the low ultraviolet
polarization, leads to the important conclusion that only the larger
grains are aspherical and aligned; the aligned grain mass distribution
is very deficient in small particles (Kim \& Martin \cite{kimm94}).



The strong 2175~\AA\ bump does not have a correspondingly dramatic
polarization feature at 2175~\AA.
A weak feature is seen only in two of the 28 lines of sight in the
Galaxy observed by WUPPE and HST (FOS), despite having sufficient S/N
(Martin \etl\ \cite{mar99}).  It is not a common phenomenon.
%
The polarization feature is at the same position as the extinction
bump, has a positive excursion, and shows no change in position angle.
But it is at least two orders of magnitude smaller than the
theoretical maximum for perfectly aligned graphite carriers (Martin
\etl\ \cite{mar95}).  Thus, either the alignment is quite incomplete
or only a small fraction of the grains is aligned.
The result that the (presumed small) particles giving rise to the
2175~\AA\ extinction bump are very poorly aligned is consistent with
the poor alignment of the (other) small particles which give rise to
the continuum ultraviolet extinction.  A small residual alignment
might be arise from the Davis-Greenstein process (Wolff \etl\
\cite{wol97}).

Significantly, the feature at 10~\micron\ is polarized
(Fig.~\ref{polcurve}, right), indicating that at least the silicate
grains are aspherical and aligned.  Details of the relative changes of
$p_{ex}$ and $\tau_{ex}$ across the feature constrain the band
strength and the shape and axial ratio; Hildebrand \& Dragovan
(\cite{hil95}) conclude that the silicate particles are oblate with
axial ratio $\sim 1.5$.  It might be relevant that this is not unlike
the individual sub-grains found in fluffy silicate agglomerate
interplanetary dust particles, and the further insight from cometary
material collected by Stardust is eagerly anticipated.

For embedded sources in molecular clouds, the polarized ice feature at
3.1~\micron\ can be interpreted as a thin coating on the aligned
silicates (as predicted: Martin \cite{mar75}).
The rise in the infrared is a power law (Martin \& Whittet
\cite{mar90}; Martin \etl\ \cite{mar92}) not unlike that for
extinction, which again can be modeled as the (differential)
scattering by large grains comparable in size to the wavelength (Kim
\& Martin \cite{kim95}).

In summary, only the larger grains are aspherical and aligned.  The
10~\micron\ polarization shows that the silicate component is aligned,
and the (oblate) particles have an axial ratio which is modest (near
1.5).  It seems at least consistent (certainly adequate) to model the
polarization with silicates alone.

\subsection{Modeling}\label{model}

The observable $p_{ex}(\nu)$ increases with the column density of
aligned grains.  It is thus useful to normalize this out by
considering the ratio with respect to extinction $\tau_{ex,\nu}$ which
also builds up with column density.  Suppressing the frequency
dependence of the cross sections,

\begin{equation}
\bigl({p \over \tau}\bigr)_{ex} = {\Sigma_g\ N_g {1\over 2}(C_{par} -
C_{per})_{ex} \over \Sigma_g\ N_g {1\over 2} (C_{par}+
C_{per})_{ex} }
\label{pot}
\end{equation}

\noindent
(see, e.g., Martin \cite{mar74}).  This has a form that is deceptively
like that for $p_{em}$, apart from the weighting of the latter with
the Planck function (see equations \ref{emit} and \ref{pemit}, and
\ref{pem}).  However, the further simplifications arising from the
approximations for $Q_{em}$ for small particles (e.g., simple volume
weighting; size-independent behaviour) do not apply.

\begin{figure}
\includegraphics[width=5cm,angle=90] {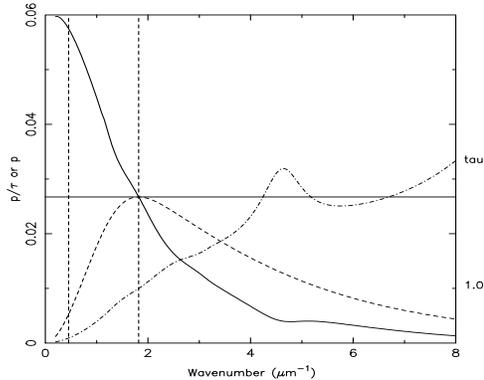}
\caption{Typical frequency dependence of $(p/\tau)_{ex}$ in the
diffuse interstellar medium (solid curve).  This is derived by
dividing a typical polarization curve (dashed; Martin \etl\
\cite{mar99}) by a typical extinction curve (dash-dot; Cardelli \etl\
\cite{car89} for $R_V = 3.1$, normalized to $\tau_V = 1$, with scale
on right).  Both $p_{ex}$ and $(p/\tau)_{ex}$ are normalized (scale on
left) according to the maximum value observed at $V$ (see
Fig.~\ref{smf}).  The ratio $(p/\tau)_{ex}$ at 2.2~\micron\ (K band,
left vertical dashed line) is significantly higher.
}
\label{pandtau}
\end{figure}

Figure~\ref{pandtau} shows the very different frequency dependences of
$\tau_{ex}$ (dash-dot) and $p_{ex}$ (dash) derived from observations
in the diffuse interstellar medium (??ccm, ??mar99).
Both $\tau_{ex}$ and $p_{ex}$ are sensitive to the actual size
distributions, for all the grains and the subset that are aligned,
respectively.  Therefore, a self-consistent model to predict
$(p/\tau)_{ex}$ and eventually $p_{em}$ requires size distributions
that reproduce the frequency dependence of both the polarization and
extinction curves.

\section{Constraining Alignment and Shape}

Both the degree of alignment and shape influence the degree of
polarization and so have to be constrained in order to predict
$p_{em}$.

\subsection{Imperfect Alignment} \label{disor}

For the ensemble of grains, disalignment can be characterized by
disorientation
of the direction of the magnetic field along the line of sight,
of the angular momentum vectors of the spinning grains with respect to
the magnetic field,
and of the spin axes (or body axes) with respect to the angular
momentum vectors (e.g., Greenberg \cite{gre68}; Purcell \& Spitzer
\cite{pur71}; Martin \cite{mar75}; Hildebrand \cite{hil88}).
%
%
These affect the emission $P$ and $I$, and $p_{ex}$ and $\tau_{ex}$,
in subtly different ways.

For grains small compared to the wavelength, as appropriate for
$p_{em}$, the cross sections for arbitrary orientations can be written
as linear combinations of the principal cross sections, so that the
average $P$ and $I$ over a distribution functions can be obtained more
straightforwardly.  For $P$ the effects are actually separable, and
not functions of shape, size, or composition, which can be useful;
Greenberg (\cite{gre68}) gives a size-independent ``Rayleigh reduction
factor''.
Of course, in the general case, the appropriate sums in $P$ over
the grain components, which might have different alignment, still have
to be performed.
In the optical there are no such approximations, but there is some
indication from representative calculations (e.g., ??green, ??rogers)
that for the same disalignment the fractional reduction in $p_{ex}$ is
quite similar to that of $P$.  Further calculations to quantify
this would be of interest, but are beyond the scope of this paper.

There are related effects through the cross sections, which are not
separable even for small particles, that tend to increase the emission
$I$ and $\tau_{ex}$ relative to perfectly aligned grains, further
reducing $p_{em}$ and $(p/\tau)_{ex}$, respectively.
The effects are small for small axial ratios, but can be more
substantial as grain flattening or elongation increases. As discussed
below, in the interstellar medium even the maximum optical
$(p/\tau)_{ex}$ corresponds to either axial ratios close to one, or
close to random orientation for more elongated particles, with some
independent evidence for the former.
Thus for $I$ and $\tau_{ex}$ calculations using spherical
particles and/or random orientation might be close to appropriate.
But without complete calculations for disaligned grains, this is one
major source of systematic uncertainty in the applying ``recipe''
below (\S~\ref{boot}).

Relative to perfect spinning alignment (PSA), which for computational
simplicity is often used, one can define a ``reduction factor'' $R$
which quantifies how the polarization is reduced when the grains are
less than perfectly aligned.
From the discussion above, $R_{ex}$ for $(p/\tau)_{ex}$ could be
different than $R_{em}$ for $p_{em}$ and be frequency dependent.
However, they are likely to be quite similar, and because little is
actually known about the details of alignment, $R_{em} \simeq R_{ex}$
is a pragmatic approximation to adopt.

\subsection{An Empirical Constraint}\label{empirical}

The observed amount of optical polarization per unit extinction
provides an empirical measure of the asphericity and degree of
alignment, whence $R$ for a particular grain model with specified
shapes and sizes.

\begin{figure}
\includegraphics[width=7cm] {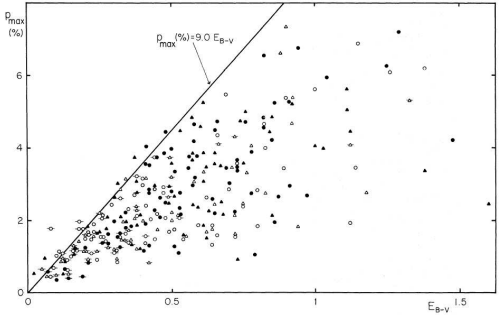}
\caption{Observed values of optical polarization near $V$ plotted
against colour excess $E_{B-V}$ (Serkowski \etl\ \cite{ser75}).  For a
typical ratio of total to selective extinction $R_V = 3.1$, the upper
envelope corresponds to $(p/\tau)_{ex} = 0.0267$ at $V$ or about 0.06
at the $K$ band (Fig.~\ref{pandtau}).  For directions in which the
magnetic field is more along the line of sight, and/or for which there
is a lower degree of alignment, $(p/\tau)_{ex}$ is lower, typically by
a factor two, and $p_{em}$ of the low frequency emission will be lower
too.
}
\label{smf}
\end{figure}

Figure~\ref{pandtau} shows $(p/\tau)_{ex}$ derived from the empirical
curves for $\tau_{ex}$ and $p_{ex}$. There is a strong systematic
spectral dependence.
The normalization is for lines of sight in the interstellar medium for
which the most optimal combination of shape and alignment is achieved
(Fig.~\ref{smf}).  This corresponds to a value $(p/\tau)_{ex} =
0.0267$ at the $V$ passband or about 0.06 at $K$.

Hildebrand \& Dragovan (\cite{hil95}) adopt a maximum, $p/\tau =
0.065$ at the $K$ band.  While this is from lines of sight with higher
extinction,and perhaps higher density, observed by Jones \etl\
(\cite{jon92}), and so perhaps less appropriate to the cirrus
application here, the value is nevertheless very similar.

\subsection{A Bootstrapping Recipe}\label{boot}

For their analysis of far-infrared polarization in more dense regions
of the interstellar medium (clouds), complicated regions for which a
maximum $p_{em} \sim 9$~\% is observed (Vaillancourt \cite{vai06};
Hildebrand \etl\ \cite{hil00}), Hildebrand \& Dragovan (\cite{hil95})
gauge the effect of disalignment by first comparing $p_{em}$ for their
model at 2.2~\micron\ with $(p/\tau)_{ex}$ observed at 2.2~\micron.
In practice, there are two problems with their approach.
First, they assumed that the formulae involving cross-sections for
pure absorption would apply to both $p_{em}$ in the far-infrared and
$(p/\tau)_{ex}$ at $K$.  However, the slope of the observed power-law
dependence of extinction and polarization near $K$ indicates that this
involves grains of size comparable to the wavelength, where scattering
is very important. The adopted formulae do not really apply at 2.2~
\micron.
Second, even if this were the case, the evaluation of $p_e$ near
2.2~\micron for silicates is very sensitive to how ``dirty'' they are,
as quantified by the size of $k$, the imaginary part of the refractive
index in the near-infrared and optical.  This is poorly known.  On the
other hand, how ``dirty'' the silicates are is not relevant to the low
frequency polarization.  Therefore, for silicates, it is hard to
bootstrap from the near-infrared to the far-infrared or submillimetre
with confidence.
Addressing the first problem with more realistic scattering models
also eliminates the second.

To proceed, as stated in \S~\ref{model} we must compute
$(p/\tau)_{ex}$ according to equation~\ref{pot} using realistic size
distributions and an electromagnetic scattering theory (Mie-like) for
particles comparable in size to the wavelength.  This is not
impossible, and so a practical bootstrapping recipe is
(i) For a given axial ratio, and perfect alignment, find the aligned
grain size distribution by fitting the frequency dependence of
interstellar polarization.  Compare this to a consistent model of the
frequency dependence of interstellar extinction, keeping track of the
mass of all components (unaligned grains contribute to $\tau$ and not
p, and so cause dilution).  For this we use models by Kim \& Martin
(\cite{kim95}); only the silicates are aspherical and aligned, which is
certainly adequate.
(ii) Calculate $(p/\tau)_{ex}$ for this model (it can be evaluated for
frequencies throughout the optical).  We took the average of the two
estimates in Table~1 of Kim \& Martin (\cite{kim95}) and noted the
implied substantial systematic error.
(iii) Compare this to the observed maximum $(p/\tau)_{ex}$
(Fig.~\ref{pandtau}) to deduce a reduction factor $R_{ex} < 1$
reflecting the imperfect alignment.
(iv) For the same model (shape, axial ratio, grain components),
self-consistently calculate the polarization of the low-frequency
thermal emission of the aligned grains, $pa_{em}$ (e.g.,
Fig.\ref{figpem}).
(v) Assuming that $R_{em} = R_{ex}$, apply $R$ from the interstellar
polarization model for that axial ratio to find the expected maximum
degree of polarization $R pa_{em}$ for the disaligned grains.

How robust this estimate is can be judged by repeating the recipe
calculation for different shapes and axial ratios.
(vi) There is a further complication, dilution of the polarization because
of thermal radiation by non-aligned grains.  This reduction factor $d
< 1$ needs to be evaluated as self-consistently as possible
(\S~\ref{dilute}).

\section{Results and Discussion}\label{results}

The results from the different steps of this recipe are collected in
Table~\ref{predictt}.  There $pa_{em}$ is evaluated at 350~GHz, but this
choice is inconsequential (Fig.~\ref{figpem}).

\begin{table}[!b] 
\caption{Prediction of the Net Submillimetre Polarization}
\begin{center}
\begin{tabular}{lccccccc}
\hline \\
                 &\multispan{4}\, Oblate\hfil &\multispan{3}  Prolate\hfil\\
Axial ratio      & $\sqrt{2}$ & 2    & 4    & 6    & 2$^1$& 2$^2$& 4$^2$\\
\hline \\
$(p/\tau)_{ex}$   & 0.04 & 0.08 & 0.12 & 0.12 & 0.05 & 0.09 & 0.14 \\
$R$               & 0.67 & 0.36 & 0.23 & 0.23 & 0.53 & 0.30 & 0.19 \\
$pa_{em}$$^3$     & 0.31 & 0.56 & 0.83 & 0.90 & 0.39 & 0.56 & 0.91 \\
$R pa_{em}$       & 0.21 & 0.20 & 0.19 & 0.21 & 0.21 & 0.17 & 0.17 \\
$d R pa_{em}$ (\%)& 9.3  & 8.9  & 8.7  & 9.4  & 9.3  & 7.5  & 7.8  \\
\hline \\
\end{tabular}
\end{center}
{$^1$ perfect spinning alignment} \\
{$^2$ picket fence alignment} \\
{$^3$ evaluated at 350~GHz} \\ 
\label{predictt}
\end{table}

For perfectly aligned oblate silicate particles ($R \sim 1$ in this
case), the axial ratio needs to be no higher than 1.4 to produce the
maximum $(p/\tau)_{ex}$ observed.  For larger axial ratios, grains
must be somewhat disaligned by a quantifiable amount ($R < 1$) to
produce the same $(p/\tau)_{ex}$.  
The value of $pa_{em}$ is large, and as expected higher for the more
extreme axial ratios (Fig.~\ref{figpem}).  The application of the
recipe to the models for axial ratios closer to unity is probably more
reliable, and these ratios are possibly the most realistic as well.
But it turns out that the product $R pa_{em}$ is fairly robust,
depending little on the shape and axial ratio.

\subsection{Dilution}\label{dilute}

Non-aligned grains contribute to the thermal emission and dilute the
maximum polarization expected.  The Kim \& Martin models used involve
silicates and (large) graphite particles (Kim \etl\ \cite{kim94}),
following Mathis \etl\ (\cite{mat77}) and Draine \& Lee
(\cite{dra84}).  The relative contributions to the submillimetre
emission can be judged from calculations by Draine \& Anderson
(\cite{dra85}), which do produce about the right amount of
far-infrared emission per unit column density.  This suggests quite a
reduction, by $d \sim 1/3$.  A more recent variant of this model, with
slightly different size distributions and apportionment of material
(Li and Draine \cite{lid01}), gives $d \sim 0.58$ at 353~GHz (their
Fig.~9).  We adopt the mean, $d \sim 0.45$, and note another major
contribution to the systematic uncertainty.

Note that the dilution of $(p/\tau)_{ex}$ in the optical by extinction
by unaligned graphite (carbonaceous) particles has resulted in a
larger $R_{ex}$, and so this low $d$ can be considered as payback,
compensating in the product $dR$.  Likewise, if there were optical
polarization by large carbonaceous particles, there would be an
accompanying decrease in $R$ but an increase in $d$, and so this
change too would not produce a greatly different net $p_{em}$ in the
submillimetre.

\subsection{Predicted Net Polarization $p_{em}$}\label{predict}

This self-consistent model then predicts a maximum net polarization $d
R pa_{em}$ as listed in Table~\ref{predictt} and summarized as the median
$p_{em} (\%) = 8.9 \pm 0.7 \pm 3.5 $.
The result is not very dependent on the shape or axial ratio of the
grains, as reflected in the modest rms 0.7\% among the different
models.
The systematic uncertainty 3.5\% 
is much more substantial, with the rough estimate based on actual
application of the recipe to the models: calculating $\tau_{ex}$
appropriately (step (ii); \S~\ref{disor}), the (frequency dependent)
dilution $d$ near 353~GHz (step (vi); \S~\ref{dilute}), and
encompassing the assumption $R_{em} = R_{ex}$ (step (v);
\S~\ref{disor}).

Thus the maximum polarization to be expected is quite large, even
without the possible boost accompanying spectral flattening of the
amorphous silicate emissivity through 350~GHz.

But when averaged over large regions with (i) non-uniform alignment
(beam dilution) or (ii) less than the optimal alignment, including the
effects changes in the direction of the magnetic field, or unfavorable
field orientation (something that changes systematically on a large
scale in the Galaxy), or (iii) alignment which changes along the line
of sight, then typically half of this might be expected, judging from
Fig.~\ref{smf} which shows the depolarization from (ii) and (iii).
This would still leave $p_{em} \sim 5$\%.

\subsection{Observations}\label{sobs}

Beno\^it \etl\ (\cite{ben04}) present Archeops measurements of
polarization in the Galactic plane at 353~GHz.  They find $p_{em} \sim
4 - 5 \%$ for the diffuse emission averaged over several square
degrees and even higher values in some large clouds.
The orientation of the $E$ vector roughly perpendicular to
the plane is reassuringly as expected.
However, no clear correlation with optical interstellar polarization
(for which the path lengths to the background stars would be shorter
than for the emission) has been established and the data are not of
high enough signal to noise to follow this at high resolution.
Because of the long lines of sight and high column densities, the
Galactic plane, even its diffuse emission, is probably not the best
place to evaluate the polarization of the higher latitude cirrus
foreground of the CMB.
The amount of polarization observed is actually close to what is
predicted by the analysis above.  Although such estimates were
actually made in 2003, before the Archeops results, this can hardly be
taken as a triumph, given the large systematic uncertainties and the
extra uncertainties of modeling the emission in the Galactic plane.

Nevertheless it does support the view that the polarization of the
cirrus at higher latitude will be significant.  Although the
variations in alignment along and across the line of sight at high
latitude are not known, it seems possible (because of the shorter
paths and thus simpler geometry being integrated) that these variations
will be less than in the Galactic plane, resulting in less
depolarization.

\subsection{Spectral Behaviour of $p_{em}$ and $P$} \label{spectp}

As noted above, the spectral index of the amorphous silicate
emissivity might be less steep than the $\beta \sim 2$ inherent in our
calculations.  A modest change was introduced by Li \& Draine
(\cite{lid01}, their eq.~1 and Fig.~9) to fit the frequency dependence
of the diffuse emission.  Boudet \etl\ (\cite{bou05}) suggest even
stronger variations with frequency over the range 150 -- 3000~GHz.
In this case, the factor $d$ would be tend to increase with decreasing
frequency as the relative importance of large silicates to the
submillimetre emission rose.  In models where it is the silicates that
produce the polarization (subscribed to here), this frequency
dependence of $d$ would result in an increase in the net $p_{em}$ at
the lower frequencies and our estimate at 353~GHz could be low by a
factor $\sim 1.6$ (not included in the systematic errors below).
It will be interesting to learn the results from B2K (and Planck)
which measure within this interesting range of frequencies.


Because of different weighting of different components, the spectral
dependence of the polarized intensity $P$ can be different than
that of total emission, $I$ (\S~\ref{emission}).
Through dilution, non-aligned grains lower the net polarization. If
the frequency dependence of emission for the diluting component is
different, then this introduces a frequency dependence into $p_{em} =
P/I$.  We have just mentioned one example.

To characterize the intrinsic emission properties of the carriers of
the polarization more directly, this confusion from dilution can be
avoided in principle by examining $P$, which would isolate the
spectral dependence of the polarizing emitters.  For frequency ranges
over which $pa_{em}$ for this component could be assumed to be
constant, this would directly give $\beta$, and its variations, for
the spectral emissivity ($I$) of this aligned-grain component.

$P$ is also the quantity on which to base statistical evaluation
of the spatial variations.

\begin{figure}[!b]
\includegraphics[width=7cm, angle=90] {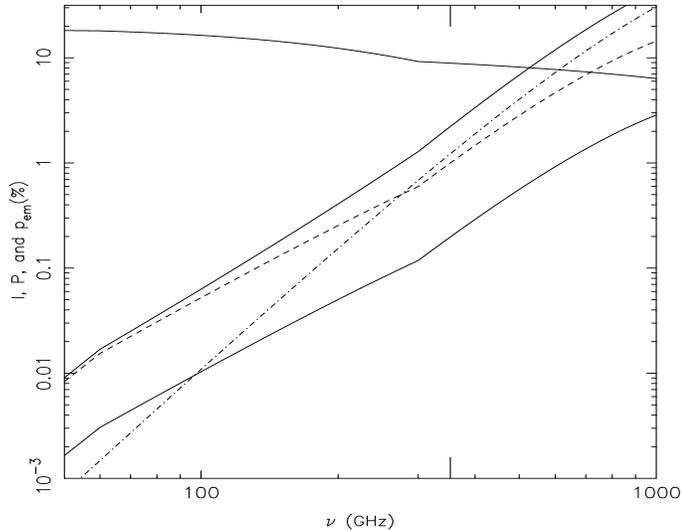}
\caption{Frequency dependence of observables $p_{em}$, $I$, and $P$
(solid curves, from top) in an illustrative model in which the value
of $\beta$ for the amorphous silicate component varies with frequency
(see text).  The silicate and graphite constributions to $I$ are shown
by the dashed and dot-dash curves, respectively (for normalization,
see text).
}
\label{toymodel}
\end{figure}

A toy model for illustration is presented in Figure~\ref{toymodel}.
Here it has been assumed that $\beta$ for the aligned silicate
component changes significantly with frequency, in the spirit of what
has been inferred from astronomical observations and measurments in
the laboratory (Boudet \etl\ \cite{bou05}; J.-P.\ Bernard, private
communication), although to be sure this is not (yet) well
constrained. Specifically the model uses
$\beta = 1.7$ for $\lambda < 1$~mm ($\nu > 300$~GHz), 
flattening (perhaps too abruptly in this illustrative model) to 0.5
through 2~mm (150~GHz), 
and then steepening to 1.5 (again for visual emphasis) beyond 5~mm
($\nu < 60$~GHz).  For the silicate grains we take $T = 17$~K and
$pa_{em}$ = 0.2 (0.089/0.45 as in \S~\ref{predict}.  In a more
realistic model there would be a slight increase in $pa_{em}$ with
decreasing frequency as a resultof impliedchanges in the complex
refractive index of the silicate.  There is also an unpolarized
graphite component with $T = 20.5$~K with relative emission such that
$d=0.45$ at 353~GHz (consistent with Table~\ref{predictt}).  For the
graphite component, $\beta = 2$.  It should be kept in mind that the
``graphite'' might alternatively be amorphous carbon, in which some
frequency dependent changes in $\beta$ might also be expected.

The ``observables'', $I$, $P$ and $p_{em}$, are shown by the solid
curves.  in Figure~\ref{toymodel}.  The model illustrates several
important features.
The spectral index measured from $I$ is a function of frequency and
would not be representative of either grain material.
On the other hand, the spectral index measured from $P$ would track
that of the aligned amorphous silicates, even though this component is
diluted in $I$.
Toward lower frequencies silicates provide a more dominant
contribution to the total emission ($d$ is larger) and so even though
$pa_{em}$ does not rise significantly for this component, the net
$p_{em}$ does increase to lower frequencies.

It appears that the frequency dependence could be quite complicated,
such that it would not be correct to extrapolate at constant $\beta$
and/or $p_{em}$ from meaurements at a single high frequency toward the
CMB frequncies.  Rather, to disentangle this most convincingly,
multi-frequency measurements of both $p_{em}$ and $P$ (and $I$) will
need to be examined in regions of relatively bright foregrounds to
produce a self-consistent model.
Such an assessment of the dust polarization on the brighter cirrus,
uncontaminated by the CMB at lower frequencies, would provide a basis
for understanding the fainter cirrus contaminating the CMB.
While this is not possible with Archeops, B2K and Planck have
polarization capability at three frequencies and PILOT will extend
this to higher frequencies (Table~\ref{angular}).
It would be advantageous to make use of polarization measurements
below 100~GHz too (e.g., WMAP and Planck), but at some point
synchrotron emission, which probably has a higher intrinsic
polarization, would no longer be negligible (Fig.~\ref{contam}), and
there is the anomalous emission to considere too..

\subsection{Alignment of Small Grains and Polarization of Anomalous Emission}\label{anomalous}

As discussed in \S~\ref{polarization}, $(p/\tau)_{ex}$ is very low in
the ultraviolet, where the extinction comes from small grains (VSGs,
PAHs).
What polarization there is (Fig.~\ref{pandtau}) is consistent with a
fading contribution coming from big grains (Kim \& Martin
\cite{kim95}).  However, it is possible that there is a very low level
of residual alignment of small grains, such as from the
Davis-Greenstein process, could be present (Kim \& Martin
\cite{kimm95}; Wolff \etl\ \cite{wol97}; Lazarian \& Finkbeiner
\cite{laz03}).

Tiny grains, including if not exclusively PAHs, are the ones that
could spin most rapidly to produce low frequency ($\sim 20$~GHz)
emission (Draine \& Lazarian \cite{dra98}), possibly accounting for
the dust-correlated anomalous microwave emission (Finkbeiner
\cite{fin04}; Finkbeiner \etl\ \cite{finl04}; de Oliveria-Costa \etl\
\cite{deo04}; Davies \cite{davies06}; Davis \cite{davis06}).
Since these small particles should not be well aligned, the degree of
polarization of that component would not be high (tiny compared to the
degree of polarization of the synchrotron component), a diagnostic
feature that might be used to advantage by WMAP and Planck LFI.
Still, the spectre of a few percent polarization for spinning dust
could be a serious contamination for precise CMB polarization
measurments (Lazarian \& Draine \cite{laz00}; Lazarian \& Finkbeiner
\cite{laz03}).

We can be somewhat more quantitative in a semi-empirical if
model-dependent way, starting from the Li \& Draine (\cite{lid01})
model, in which PAHs are responsible for most if of the strong
2175~\AA\ feature in the extinction curve (as reasonable and
economical assumption).  We then use the tight limits placed on the
polarization of the 2175~\AA\ feature (Martin \etl\ \cite{mar99};
\S~\ref{polarization}).  The relative change in polarization, compared
to extinction, at the bump gives for this component $p/\tau_{ex} <
0.002$.  The same quantity for perfectly aligned PAHs could be
calculated using small-particle formulae as in equations
\ref{qspheroid} and \ref{pem} (the extinction is pure absorption),
given a shape and refractive indices.  However, there are no
refractive indices since only absorptivity has been modeled (Li \&
Draine \cite{lid01}), and for these large molecules it might not even
be particularly appropriate.  Nevertheless, since the PAHs are
probably highly flattened, this quantity would be near unity, just as
it is for tiny graphite spheroids of even modest asphericity (Martin
\etl\ \cite{mar95}; Wolff \etl\ \cite{wol97}).  Therefore, $R_{ex} <
0.002$.  Similarly, for the rotational dipole emission of well aligned
tiny grains, $pa_{em}$ would also be unity.  It would be reasonable to
assume similar effects from disorientation: $R_{em} \sim R_{ex}$.
Thus for the dust-related anomalous emission, we would predict a
degree of polarization less than 0.2\%, probably very much less given
that polarization of the 2175~\AA\ feature is unusual.  This is more
than an order of magnitude, probably nearer two, less than for the
degree of polarization of the thermal dust emission, and also very
much less than expected for synchrotron emission.  The real situation
is more subtle, however, because polarization of the rotational
emission depends on the alignment of the angular momentum vector with
respect to the magnetic field rather than the body axis with respect
to the field, needed for the ultraviolet polarization (Lazarian \&
Draine \cite{laz00}). If there is only partial alignment of the
angular momentum vector and the body axis, then the low frequency
polarization would be higher and with detection of polarization of the
anomalous emission one might learn something more about the alignment.

To complicate things further, there is an alternative model for the
anomalous microwave emission, magnetic-dipole emission from magnetic
grains (Draine \& Lazarian \cite{dra99}; Lazarian \& Finkbeiner
\cite{laz03}), which would have a very distinctive frequency-dependent
polarization signature.  Not enough is known about the properties of
these grains or their alignment to make firm predictions of the degree
of polarization.  However, since magnetic inclusions might be a factor
in grain alignment, this emission might be more closely related to
``normal-sized'' grains which are better aligned, and therefore might
be more highly polarized than emission from spinning dust.  Following
similar bootstrapping arguments such as above, the order of magnitude
might be like that given by $R pa_{em}$ in Table~\ref{predictt}, thus
$\sim 20$\%, reduced by other diluting contributions to the microwave
emission.  Again, synchrotron polarization would be a consideration.

\acknowledgements This work was supported by the Natural Sciences and
Engineering Research Council of Canada. I thank F.\ Boulanger, J.-L.\
Puget and colleagues for hospitality at IAS in 2003 where much of this
work was done and F.\ Boulanger and M.-A. Miville-Desch\^enes
for encouraging me to publish the results!


\noindent
This volume is {\it Sky Polarisation at Far-infrared to Radio
Wavelengths: The Galactic Screen before the Cosmic Microwave
Background}, {eds.\ M.-A. Miville-Desch\^enes and F. Boulanger. }{EAS
Publications Series, Vol. --: Paris}

\vspace{0.1 in}


\noindent
{\bf Appendix: The Last Word -- Grain Alignment Explained}

\vspace{0.05 in}

\noindent
I recommend looking at the above-mentioned Planck web pages, but one
thing that cannot be found there is an explanation of why the grains
are aligned.  Recall that somehow the dust particles are supposed to
be spinning preferentially around the magnetic field.
At the conference I announced my forthcoming monograph ``Alignment for
Dummies'' (coming soon to a discerning supermarket checkout counter
near you) and to pique curiosity further (advanced sales benefit from
feverish anticipation) I elaborated on the merits of one new theory of
grain alignment.
It turns out that the new theory is rather straightforward, so much so
that I attempted to explain it {\it en fran\c cais}, hoping to
motivate, indeed enable, Lagache and colleagues to update their Planck
web pages.

\vspace{0.05 in}

{\it
\noindent
C'est facile, comme ``un, deux, trois.'' 

\noindent
Un! Voila, une grande poussi\`ere.  Je pense que c'est organique, \`a la
Greenberg.

\noindent
Deux! Voila, le champs magn\`etique interstellaire -- plus grand que les
Champs-Elys\'ee.

\noindent
Mais, aucun alignement!  Dommage!

\noindent
Trois! Voila, la main de Dieu!   

\noindent
Alors, attention!  Un: une poussi\`ere.  Deux: un champs magn\`etique.
Trois: la main de Dieu.  Alignement!
}

\vspace{0.05 in}

\begin{figure}
\includegraphics[width=7cm] {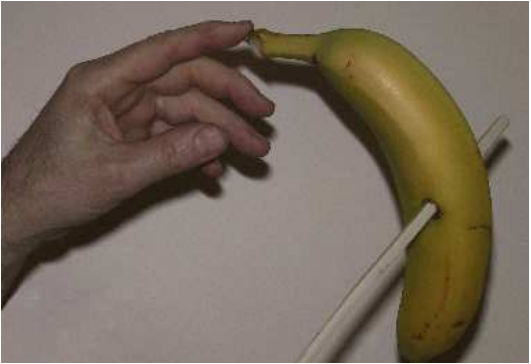}
\caption{A moment of great anticipation in the demonstration of
``Intelligent Alignment,'' at the point of ``contact'' as the
propeller motion is energized.
}
\label{maindedieu}
\end{figure}

The live demonstration (Fig.~\ref{maindedieu}) does benefit from a
grocery bag containing appropriate ingredients!
Any good theory needs a name and it is suggested that this theory
might well merit the status of ``Intelligent Alignment.''

\end{document}